\documentclass[12pt,a4paper,fleqn]{scrartcl}

\usepackage{graphicx} 
\usepackage{amsmath} 
\usepackage{amssymb} 
\usepackage{hyperref} 
\usepackage{authblk} 
\usepackage{natbib} 

\usepackage{framed}
\usepackage{lipsum}
\usepackage{verbatim}
\usepackage{authblk}

\title{A Modern Paradigm for Algorithmic Trading}

\author{J.B. Glattfelder, T. Houweling, and R.B. Olsen}
\date{\small \today}

\begin{document}

\maketitle

\begin{abstract} \noindent
We introduce a novel framework for developing fully-automated trading model algorithms. Unlike the traditional approach, which is grounded in analytical complexity favored by most quantitative analysts, we propose a paradigm shift that embraces real-world complexity. This approach leverages key concepts relating to self-organization, emergence, complex systems theory, scaling laws, and utilizes an event-based reframing of time. In closing, we describe an example algorithm that incorporates the outlined elements, called the Delta Engine.
\end{abstract}

\tableofcontents

\section{Introduction}

Following the Pythagorean-Aristotelian tradition, mathematics was always understood to be the fundamental language for decoding the structure and function of the universe. This promise reached its full potential about 350 years ago with the rise of the modern scientific age, initiated by Isaac Newton and Gottfried Wilhelm Leibniz's introduction of calculus. The subsequent mathematization of reality has, to this day, unlocked breathtaking technological prowess. However, one domain continues to elude the inquiry of the analytical human mind. The complexity we witness around us---and within us---appears to defy mathematical description \citep{anderson1972more}. For example, the emergent phenomena of life and consciousness remain enigmatic \citep{glattfelder2025sapient}.

To address the hallmarks of complexity---characterized by non-linear, path-dependent, fractal, and chaotic dynamics influenced by feedback mechanisms \citep{holland2014complexity}---complexity science emerged as a fusion of many intellectual traditions. Today, complex systems theory builds upon a rich legacy, including cybernetics, systems theory, non-equilibrium thermodynamics, agent-based modeling, and network science.

Financial markets are the epitome of complexity, arising from a worldwide network of interacting agents and representing the backbone of our globalized economy. Interestingly, ideas from complexity theory do not seem to be prevalent in quantitative finance, mostly assuming linearity, equilibrium states, and utility functions. Mathematical elegance often takes precedence over data-driven analysis \citep{glattfelder2019applied}.

Especially the development of algorithmic trading strategies appears to be firmly embedded within the analytical paradigm. Nonetheless, devising fully automated, consistently profitable strategies remains an elusive challenge. Countless clever minds have failed to meet this challenge. Either automated trading is an unattainable fantasy or researchers are operating from within the wrong paradigm. Efforts to devise trading model algorithms should embrace complexity at its core, leveraging principles of self-organization, adaptivity, and resilience to create trading strategies that align with the true nature of financial markets. In the following, the core building blocks of the framework are described.

\section{Complex Systems}

It is a remarkable fact that the formal thought systems that the human mind can devise have the capacity to gauge the workings of many natural phenomena. Mathematics is the machine code of reality and a dialect the rational mind is naturally conversant in \citep{wigner1960unreasonable}. In the domain of complexity, an additional surprising feature of reality emerged, allowing complexity to be tamed. What appears as complex behavior from afar often arises from simple rules governing local interactions \citep{wolfram2002new,wolfram2020project}, a process that lies at the heart of emergence. This is a phenomenon representing a discrete quantum leap in organization, where a system as a whole displays novel properties that cannot be deduced solely from analyzing its parts in isolation. Only by simulating the system's evolution as a whole do the self-organizing dynamics become visible. As a result, an algorithmic approach to decoding complexity is an essential tool, revealing structures and patterns that traditional analytical methods overlook.

In summary, operating within the mathematical paradigm, the equations describing a natural system can be solved to provide insights into the nature of the phenomena under investigation. For complex systems, however, the chosen rules of interaction determine the system's behavior. Applying this insight to the context of trading model algorithms, the emphasis thus shifts from equation-solving to an agent-based approach simulating the intrinsic dynamics---a true paradigm shift in understanding.

While complex systems can be encountered at multiple scales, from micro to cosmic, and across various domains, they seem to adhere to one specific type of organizing principle. Functioning much like a law of nature, this feature guides the dynamics of the system and results in surprising regularities.

\section{Scaling Laws}

A scaling law describes a simple polynomial relationship in which a change in input results in a proportional change in output
\begin{equation*}
   f(x) = C x^\alpha, 
\end{equation*}
where $\alpha$ is a positive or negative scaling parameter and $C$ a constant.

Scaling laws exhibit scale-invariant behavior, meaning they lack a preferred or defining scale\footnote{Formally, $f(ax) \sim f(x)$.}---a fundamental characteristic of fractals. Such laws are ubiquitous, appearing across diverse domains, including physics, biology, Earth and planetary sciences, computer science, demography, social sciences, economics, and finance \citep{newman2005power}.

In financial time series, scaling laws relate the behavior observed at small scales to the behavior at larger scales, revealing consistent patterns across different magnitudes \citep{glattfelder2011patterns}. For example, price fluctuations over short time intervals may exhibit similar statistical properties as those observed over longer intervals. As a result, measuring the behavior of financial time series at only two resolutions allows the scaling-law relation to be estimated, unveiling a wealth of information from an unexpected and robust statistical property.

\section{Intrinsic Time}

The notion of time mostly plays a minor role in finance and economics. It is assumed that time flows continuously and that time series data should be analyzed at regular, equidistant intervals. However, an event-based measure of time appears more appropriate for dealing with the complex behavior seen in financial markets. Such a conception of time accelerates during periods of high activity and slows down when conditions are quieter, capturing the true rhythm of markets.

One special implementation of an event-based measure of time is called intrinsic time \citep{dacorogna1996changing}. It is defined by two atoms of activity observed in time series: directional changes and overshoots \citep{glattfelder2024theory}. Given a percentage threshold $\delta$, a directional change occurs when the price evolution reverses by $\delta$ (from a local extrema). After such a reversal the price can continue in the same direction in increments of $\delta$, characterizing the overshoots. Intrinsic time ticks when directional changes or overshoots are measured. By construction, it is a multi-scale framework, allowing observations at various granularities. Moreover, the framework can easily be expressed algorithmically. It functions as an operator that takes the original time series as an input and produces a new one solely comprised of intrinsic time events at a predefined threshold.

Many scaling-law relations emerge within the framework of intrinsic time, expressing signs of self-organizing complexity. As an example, two salient scaling laws uncover hallmark properties in market data: the number of directional changes \citep{guillaume1997bird} and the average overshoot length \citep{glattfelder2011patterns}. While the former can be utilized to define a proxy for volatility \citep{petrov2019instantaneous}, the latter allows a liquidity measure to be constructed \citep{golub2016multi}. It can be shown that the number of directional changes times the variability of overshoots is equal to the squared returns sampled equidistantly in physical time, thus linking the two temporal frameworks \citep{glattfelder2022bridging}.

\section{The Delta Engine: An Unorthodox Approach to Algorithmic Trading}

To illustrate the principles outlined in this paper, we present a trading model algorithm that aligns with the described paradigm. The \textit{Delta Engine} utilizes the following ingredients:
\begin{itemize}
    \item Intrinsic time reduces time series to their essential atoms of activity across multiple scales $\{\delta_1, \dots, \delta_n\}$, thus operating at different physical time horizons. The algorithm is implemented as an agent-based framework, where each $\delta_i$ threshold gives the agent's context.
    \item On these transformed data landscapes, the trading model builds the framework for adaptive decision-making: For each threshold and various look-back sizes, resistance and support lines are fitted to the upward and downward overshoot events, respectively.
    \item A breakout signal is triggered if the price evolution of a threshold breaches a support or resistance line.
    \item These multi-scale signals are aggregated and trigger trades only if one of the thresholds detects a contrarian pattern, enforcing emergent collective decision-making.
    \item A feedback loop is realized by accounting for the market activity. Specifically, the number of directional changes scaling law acts as a volatility measure reflecting the current market state. Calibrating to this rhythm, agents can fall out of sync and be temporarily silenced, further enforcing self-organized decision making. In other words, the volatility regime directly fine-tunes the model behavior.
\end{itemize}

The Delta Engine is unique in that it exclusively trades on contrarian breakout signals. When an initial signal occurs, the algorithm executes a long or short trade of size $u$. Now, only a breakout signal identifying a reversal of the initial trend will generate an offsetting trade of size $2 u$. As a result, the net exposure oscillates between $+u$ and $-u$. Essentially, the Delta Engine enforces a simple dichotomy resulting from alternating signals indicating the opening of positions. Unlike traditional trading strategies which implement predefined take-profit and stop-loss thresholds, the intrinsic behavior of the Delta Engine is decoupled from the PnL evolution. Thus, a signal always represents a moment in time when the multi-scale behavior of a currency pair holds predictive power over how the future will unfold.

It should be noted that the algorithm's behavior is determined by only a few configuration parameters. By design, the set of intrinsic time thresholds is quintessential. Then, the fitting of the trend lines requires additional configuration choices, as does the volatility proxy. Unlike conventional trading algorithms burdened by an intractable parameter space, the Delta Engine adopts a minimalist approach to configurability, effectively reducing the risk of overfitting. At its core, it is characterized by self-organizing behavior. 

In summary, the Delta Engine's trading behavior is fully emergent, arising from the intrinsic states of its interacting agents given the current market dynamics. It thus embodies the hallmarks of complexity, expressing adaptive and resilient dynamics that arise from simple rules of interaction. The presented paradigm has a rich structure---and a long history \citep{glattfelder2014r}---offering unique insights and innovative approaches that challenge conventional methodologies, laying the groundwork for novel understanding and applications in algorithmic trading. In future work, more details will be unveiled.


\bibliographystyle{apalike} 
\bibliography{example} 

\end{document}